\documentclass[aps,prl,superscriptaddress,twocolumn]{revtex4}

\usepackage{graphicx} 
\usepackage{amsmath}
 
\newcommand{\beq}{\begin{equation}}
\newcommand{\enq}{\end{equation}}

\begin{document}

\title{Quantum fluctuations of a vortex in an optical lattice}
\author{J.-P. Martikainen}
\author{H. T. C. Stoof}
\affiliation{Institute for Theoretical Physics, Utrecht University, 
Leuvenlaan 4, 3584 CE Utrecht, The Netherlands}
\date{\today}

\begin{abstract}
Using a variational ansatz for the wave function of the Bose-Einstein condensate,
we develop a quantum theory of vortices and quadrupole modes in a one-dimensional
optical lattice. We study the coupling between the
quadrupole modes 
and Kelvin modes, which turns out to be formally analogous to the
theory of parametric processes in quantum optics. This leads to the possibility
of squeezing vortices. We solve  
the quantum multimode problem for the Kelvin modes and quadrupole modes 
numerically and find properties that cannot be explained with a simple 
linear-response theory.
\end{abstract}
\maketitle

{\it Introduction.}---
There is a large body of research on vortices in Bose-Einstein condensates,
both before and after their first experimental 
observation~\cite{Matthews1999b,Madison2000a}.
A recent review of this vibrant field was given by Fetter and
Svidzinsky~\cite{Fetter2001a}. One reason for this interest is that
vortices play a crucial role in explaining
rotational and dissipative properties of superfluids.
They are also important for an understanding of classical fluids and 
wave excitations of a vortex 
were therefore already studied by Lord Kelvin. 
Such excitations have been observed in superfluid helium~\cite{Hall1958a}
and, most recently, in Bose-Einstein 
condensates by Bretin {\it et al.}~\cite{Bretin2003a}. The latter experiment
also elucidated the role of the quadrupole modes of a Bose-Einstein condensate
in exciting the Kelvin modes of the vortex line. A theoretical understanding
of the observed phenomena 
was provided by Mizushima {\it et al.}~\cite{Mizushima2003a}.

The phenomena of superfluidity is closely connected to the existence
of quantized vortices and relies on
phase coherence across the condensate. Placing the condensate in an optical lattice makes 
the establishment of phase coherence more difficult and can even induce a transition into 
a Mott-insulator state. This quantum phase transition was indeed recently observed
experimentally~\cite{Greiner2002a}. 
In a one-dimensional optical lattice the condensate splits into a stack of weakly-coupled 
disk-shaped condensates, which leads to some intriguing 
analogs with high-$T_c$ superconductors due to their similar
layered structure~\cite{Blatter1994a}. 

In this Letter we combine the above ideas and
study a vortex in a Bose-Einstein condensate in a one-dimensional 
optical lattice.
This problem has not yet been addressed either experimentally or theoretically, but
experimental studies are possible with current technical capabilities.
Indeed such an experimental setup appears to be a promising way
to achieve the quantum Hall regime in a Bose gas, for which we need
to have about one vortex per particle~\cite{Cornell2002private}.
Also outside the quantum Hall regime an optical lattice allows for 
unique possibilities 
to study vortices and their quantum fluctuations.
In a normal three-dimensional condensate a vortex 
line is a very complicated dynamical object and it is nearly impossible
to tune its properties in a controlled way. 
Placing the vortex in an optical lattice provides a way, for example, 
to reduce the stiffness of the
the vortex line by increasing the depth of the lattice. 

With this goal in mind, 
we develop a quantum theory for the Kelvin modes in an optical lattice and
obtain the Kelvin mode dispersion relation
analytically by means of a variational approach.
Furthermore, because of their importance for the creation of the Kelvin waves,
we study in detail the coupling between the Kelvin modes
and quadrupole modes.
Our approach leads to a quantum multimode
problem with an interaction between Kelvin modes and quadrupole modes
that closely resembles squeezing Hamiltonians familiar from quantum optics.
We solve this quantum multimode problem 
and show how the squeezing is reflected
in the quantum mechanical uncertainty of the vortex line. 
In addition, numerical solution of the multimode problem
reveals properties of the nonlinear dynamics that cannot be
explained with a linear-response theory. 

{\it Theory.}---
The theory we use is based in our previous work~\cite{vanOosten2001a,Martikainen2003a}, 
and we briefly summarize the
main points.
We consider an elongated Bose-Einstein condensate in a magnetic trap 
with a radial trapping frequency $\omega_r$ that is much larger than
the axial trapping frequency $\omega_z$.
In addition to this cigar-shaped potential
the condensate experiences an one-dimensional optical lattice 
$V_0\left({\bf r}\right)=V_0\sin^2\left(2\pi z/\lambda\right)$,
where $V_0$ is the lattice depth and $\lambda$ is the wavelength of the 
laser light. Furthermore, we assume that the lattice is deep enough so that it
dominates over the magnetic trapping potential in the $z$ direction and
that the number of sites are very large. Then we can in first instance 
ignore the magnetic trap in the $z$ direction.

We expand the condensate wave function in terms of wave functions 
$\Phi_n\left(x,y\right)\Phi(z)$
that are well localized in the sites labeled by $n$.
We do not yet assume anything about the wave functions $\Phi_n\left(x,y\right)$ for the
two-dimensional condensates, but for the wave function
in the $z$ direction, $\Phi\left(z\right)$, we use the ground-state
wave function of the harmonic approximation to the lattice potential near the
lattice minimum. This harmonic trap has the frequency 
$\omega_L=\left(2\pi/\lambda\right)\sqrt{2V_0/m}$, where $m$ is the atomic mass.
These considerations lead to an energy functional of the 
weakly-coupled two-dimensional  condensates, which is characterized by the
strength of the on-site particle interactions 
$4\pi U=4\pi N\left(a/l_r\right)\sqrt{\omega_L/2\pi\omega_r}$
and by the strength of the nearest-neighbor Josephson coupling 
$J=\left(1/8\pi^2\right)\left(\omega_L/\omega_r\right)^2
\left(\lambda/l_r\right)^2\left[\pi^2/4-1\right]
e^{-\left(\lambda/4\,l_L\right)^2}$~\cite{Martikainen2003a}. In these formulae 
$N$ is the number of atoms in each site, $a$ is the scattering length, and 
$l_L=\sqrt{\hbar/m\omega_L}$.
Also, we use trap units so that the 
unit of energy is $\hbar\omega_r$, the unit of time is $1/\omega_r$,
and the unit of length is $l_r=\sqrt{\hbar/m\omega_r}$.
Our theory is valid when the lattice depth is much bigger than
the chemical potential. Also, the strength of the Josephson coupling should
be small, but still large enough to support a superfluid state. The superfluid 
Mott-insulator
transition in one-dimensional optical lattice is expected to occur only for very deep
lattices~\cite{vanOosten2003a}, and such a transition has indeed not yet been observed
experimentally.

{\it Variational ansatz.}---
A variational ansatz for the two-dimensional  condensates
should satisfy two obvious physical criteria. First,
the ansatz must predict the vortex precession dynamics correctly. 
Second, the ansatz must predict the quadrupole mode frequencies with sufficient accuracy.
In an earlier publication 
we used an ansatz that fulfilled the second requirement very well in the full
parameter range from the
non-interacting limit to the Thomas-Fermi regime~\cite{Martikainen2003a}. 
Unfortunately, this ansatz
does not fulfill
the first requirement because it does not properly account 
for the size of the vortex core, which is not important for the collective modes
but plays a crucial role in the dynamics of the vortex.
Therefore, we use the ansatz 
\begin{widetext}
\beq
\Phi_n\left(x,y\right)\propto \exp\left(-\frac{B_0}{2}\left(x^2+y^2\right)
-\frac{\epsilon_n(t)}{2}\left(x^2-y^2\right)-\epsilon_{xy,n}(t)xy
+i\tan^{-1}\left(\frac{y-y_n(t)}{x-x_n(t)}\right)\right),
\label{ansatz}
\enq
\end{widetext}
where $(x_n,y_n)$ is the position of the vortex in the $n$th site and the
variational parameters $\epsilon_n$ and  $\epsilon_{xy,n}$ are complex. 
In addition, the square of the wave function is normalized to $N$. We have chosen our
variational parameters such that only Kelvin modes and quadrupole modes are included.
Our theory can be easily adapted to include the monopole
mode as well, but we choose to focus on the quadrupole modes 
due to their
experimental importance for the creation of Kelvin waves~\cite{Bretin2003a}.
Since our ansatz does not have a core, the average kinetic energy diverges.
Therefore, we introduce a small distance cut-off $\xi$ as an additional
variational parameter. Together with $B_0$ this parameter will be determined
by minimizing the equilibrium energy functional. This minimization
is a simple numerical task that can be done once in the beginning of
the calculation. In that manner the condensate with a vortex is slightly
bigger than a condensate without a vortex and the vortex core size becomes
smaller with increasing atom numbers. Quantitatively, we find 
to a very good accuracy that $\xi=\sqrt{B_0}$.

The ansatz in Eq.~(\ref{ansatz}) fulfills both our requirements with physically
realistic parameter values and is still simple enough to be analytically 
tractable. Our ansatz fails for very weakly-interacting condensates, but
provides an accurate description when the interaction strength $U$
is larger than about $10$. In the other regime our previous ansatz 
can be used~\cite{Martikainen2003a}. In this Letter we do not consider
rotation of the trap, but
a change to the  rotating frame can again be easily included into our approach.

\begin{figure}[bht]
\includegraphics[width=0.8\columnwidth]{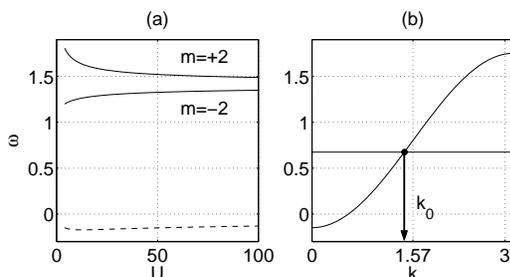}
\vspace*{-0.5cm}
\caption[fig1]{(a) Frequencies of the $m=\pm 2$ quadrupole modes 
and the Kelvin mode mode at $k=0$ as a function of interaction strength.
(b) The constant line is $\omega_{-2}\left(0\right)/2 $ and the curve is the Kelvin mode dispersion.
In linear response the energy transfer between the $m=-2$ quadrupole modes and the 
Kelvin modes becomes nonvanishing if the Kelvin mode frequency matches $\omega_{-2}\left(0\right)/2$,
i.e., when the lines cross. We used $U=100$ and $J=0.1$.
\label{omegas}}
\end{figure}

{\it Quantization of the vortex and the quadrupole modes.}---
If we expand the Lagrangian for our system to second order
in the deviations from equilibrium, we are in a position
to study the eigenmodes of the system. The eigenmodes we find are the Kelvin modes
and two quadrupole modes with quantum numbers $m=\pm 2$. To second order
there is no coupling between the vortex positions and the quadrupole modes and
we can treat the two independently.
We transform to momentum space using the convention
$f_k=1/\sqrt{N_s}\sum_n e^{-ikz_n} f_n$, where $N_s$ is the number of sites.
We then proceed by defining the vortex positions in terms
of bosonic creation and annihilation operators as 
$\hat{x}_k=(\hat{v}_{-k}^\dagger+\hat{v}_k)/2\sqrt{NB_0}$ and 
$\hat{y}_k=i(\hat{v}_{-k}^\dagger-\hat{v}_k)/2\sqrt{NB_0}$~\cite{Fetter1967a}. 
With this definition the Lagrangian
for the vortex position takes the desired form 
\beq
\hat{L}_K=\sum_k \left(i\hat{v}_{k}^\dagger\dot{\hat{v}}_k-
\omega_K\left(k\right)\hat{v}_{k}^\dagger\hat{v}_k\right),
\enq
where, to leading order in $B_0$, the Kelvin mode dispersion is given by
$\omega_K(k)=B_0/2+\Gamma\left[0,B_0^2\right]\left(2J(k)-B_0/2\right)$. In this
expression $J(k)=J\left(1-\cos\left(kd\right)\right)$, $d=\lambda/2$, and
$\Gamma\left[a,z\right]$ is the incomplete gamma function.
Our result for the Kelvin mode dispersion relation has some interesting 
features as seen in Fig.~\ref{omegas} (b). First, the initially negative precession frequency,
that indicates that a straight vortex line is unstable without rotation,
can change sign
if the momentum $k$ of the Kelvin mode is large enough. 
Moreover, the dispersion is quadratic for small momenta. Both features are due to
the attractive interactions between neighboring pancake vortices, which is harmonic
for small separations and due to 
the energy cost for having phase differences between sites.
The quadratic Kelvin dispersion in a lattice 
contrasts with the behavior of the vortex line in a three-dimensional
bulk superfluid, where the Kelvin mode dispersion 
relation behaves like $k^2 \ln\left(1/k\xi\right)$ and
has a logarithmic dependence for small $k$. 
Third, the vortex position is ``smeared'' by quantum fluctuations. 
In particular we have for the Kelvin mode vacuum state that 
$\langle \hat{x}_n^2\rangle=\langle \hat{y}_n^2\rangle=1/\left(4NB_0\right)$. 
Therefore, the quantum properties of the vortex 
become more important in a lattice. This is due to
the reduced on-site particle number, as opposed to the total number of particles,
and the spreading out of the condensate wave function
as the lattice depth is increased.

The quadrupole modes are technically somewhat more complicated, but can be quantized in a 
similar way as the Kelvin modes. The quadrupole mode frequencies are
\begin{widetext}
\vspace{-0.5cm}
\beq
\omega_{\pm 2}\left(k\right)=\frac{B_0}{A\left(k\right)}+
A\left(k\right)\left(\frac{1}{B_0}+\frac{B_0}{4}
-UB_0-\frac{B_0\Gamma\left[0,B_0^2\right]}{2}\right)+
2J(k)\left(A\left(k\right)+\frac{1}{A\left(k\right)}\right)
\pm B_0,
\enq
\end{widetext}
where
\beq
A\left(k\right)=\sqrt{\frac{4B_0^2+8B_0J(k)}{4+8B_0J(k)+
B_0^2\left(1-4U-2\Gamma\left[0,B_0^2\right]\right)}}.
\enq
The Lagrangian for the quadrupole modes is then
\beq
\hat{L}_Q=\sum_{m=\pm 2}\sum_k \left(i\hat{q}_{m,k}^\dagger\dot{\hat{q}}_{m,k}
-\omega_m\left(k\right)\hat{q}_{m,k}^\dagger\hat{q}_{m,k}\right).
\enq
In Fig.~\ref{omegas} (a) we plot the behavior of the mode frequencies at $k=0$
as a function of the interaction strength. The quadrupole mode frequencies are 
in good agreement with the ones calculated using the 
Bogoliubov-de Gennes equation~\cite{Martikainen2003a}
and with realistic interaction strengths the vortex precession frequency is 
close to the result
using a Thomas-Fermi wave-function for the disk-shaped condensate~\cite{Fetter2001a}.
\vspace{-0.3cm}

\vspace{-0.2cm}
{\it Coupled Kelvin modes and quadrupole modes.}---
We now proceed to apply our previous results to study 
the coupling between the Kelvin modes and the quadrupole modes.
To obtain coupling between the quadrupole modes and the Kelvin modes, we
must expand the Lagrangian up to third order. 
At third order there
are contributions from the kinetic energy, the time-derivative term in the 
Lagrangian, and the Josephson coupling term. Since
we assume small $J$ we can safely ignore the last one. 
The remaining terms describe the coupling between the Kelvin modes and
both quadrupole modes.
The contribution from the time-derivative term in the Lagrangian includes
the time-derivative of the Kelvin mode operators. In the rotating-wave approximation
we can replace these time-derivatives with 
the Kelvin mode dispersion. In addition,
this eliminates the nonresonant terms from the Hamiltonian, which in particular 
removes the coupling between the Kelvin modes and the $m=2$ quadrupole mode. 
In this way we obtain the following interaction Hamiltonian for the Kelvin modes and the
quadrupole modes
\vspace{-0.1cm}
\beq
\label{interactionmm2}
\hat{H}_I=\sum_{k,k'} E_c\left(k,k'\right)
\left(\hat{q}_{-2,k'}\hat{v}_{k+k'}^\dagger\hat{v}_{-k}^\dagger 
+\hat{q}_{-2,k'}^\dagger\hat{v}_{-k-k'}\hat{v}_{k}
\right), 
\enq
where 
$E_c\left(k,k'\right)=
\sqrt{A(k')/8N_s}\left(B_0\Gamma\left[0,B_0^2\right]+
\omega_K\left(k\right)\right)$.
This expression is accurate as long as the
coupling between the Kelvin modes and the quadrupole modes is sharply peaked
around some resonant value $k_0$. Note that the angular momentum is conserved, since the
quadrupole mode has an angular momentum of $-2$ whereas each Kelvin mode has an
angular momentum of $-1$.

{\it Multimode problem.}---
Our interaction Hamiltonian shows how the coupling singles out a wavenumber
for the Kelvin modes and thus selects the wavelength for the ensuing wiggles
in the vortex line~\cite{Mizushima2003a}. 
We demonstrate this in Fig.~\ref{omegas} (b).
The width of the resonance is still unknown, but we can
give a simple estimate. Assume that initially we have a straight vortex line
and a coherent quadrupole field $\langle \hat{q}_{-2,k}(0)\rangle=q_{-2,0}(0)\delta_{k,0}$. 
At short times we can assume
that the quadrupole field stays constant and then solve the operator equation
for the Kelvin modes $\hat{v}_{\pm k}$. It turns out that the solutions are dynamically 
unstable only when $|\Delta\omega\left(k\right)|<
|E_c\left(k,0\right)q_{-2,0}(0)|$, where
 $\Delta\omega(k)=\omega_{-2}\left(0\right)-2\omega_K\left(k\right)$.
Our  numerical solutions of the multimode
problem confirm this simple estimate with a reasonable accuracy. 
The number of modes within the interval 
depends on the number of sites in the lattice and on
the quadrupole amplitude, but with realistic
parameter values the multimode behavior is more likely than a single-mode
behavior. 

\begin{figure}[h]  
\vspace{-0.3cm}
\includegraphics[width=\columnwidth]{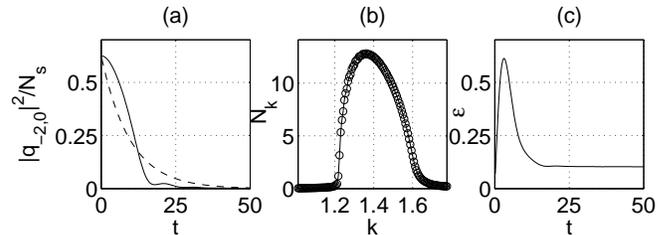}
\vspace*{-0.5cm}
\caption[fig2]{(a) The time-evolution of the quadrupole mode (solid line)
compared with the result based on Fermi's golden rule (dashed line).
(b) The number distribution of the Kelvin modes at the end of the simulation,
when the quadrupole amplitude has essentially vanished. The circles indicate the
individual Kelvin modes included in the simulation and the solid line
is an interpolation between the data points.
(c)
The time-evolution of the asymmetry factor $\epsilon\left(t\right)$.
We used $U=100$, $J=0.1$, $N_s=1001$, $N=100$, and $\langle \hat{q}_{-2,0}(0)\rangle=25$. 
\label{Dynamicsplot}}
\end{figure}

In linear response we can use Fermi's golden rule to calculate the transition rate from
the initially coherent state 
$|\psi_i\rangle=|q_{-2,0},\left\{v_k\right\}\rangle=|q_{-2,0},\left\{0\right\}\rangle$.
After a short calculation we get the transition rate as
\beq
\Gamma_{Q\rightarrow K}=
\frac{\pi E_c(k_0,0)^2|q_{-2,0}\left(0\right)|^2}{2J\Gamma[0,B_0^2]|\sin\left(k_0d\right)|}.
\enq
In Fig.~\ref{Dynamicsplot} we show an example of the time evolution
of the system, when the quadrupole field is treated as a classical field, but
Kelvin modes are treated fully quantum mechanically.
In the same figure, we also compare the exponential 
$q_{-2,0}\left(0\right)e^{-\Gamma_{Q\rightarrow K} t}$
with the numerical solution
of the nonlinear multimode problem and show the number distribution of Kelvin modes
at long times. It should be noted that at these time scales 
the system appears irreversible due to the large number of modes present. If
the number of sites in the lattice is smaller the oscillatory behavior becomes 
stronger. It should also be realized that, depending on parameters,
the result from Fermi's golden rule
can be  unreliable since it ignores the finite lifetime of the final states.
It is interesting to observe that the number distribution of the
Kelvin modes is asymmetric. This is due to the asymmetric nature of the coupling
constant $E_c\left(k,0\right)$ and the detuning $\Delta\omega\left(k\right)$.

The interaction Hamiltonian in Eq.~(\ref{interactionmm2}) 
closely resembles the squeezing Hamiltonian encountered for parametric processes
in quantum-optics~\cite{Scully1997a}. Therefore, it is not surprising 
that we can observe a dramatic squeezing under certain conditions.
For example, if we only have a single resonant mode
then the initial minimum uncertainty state, i.e., the vacuum, will remain a minimum
uncertainty state, but can become strongly squeezed. Such a squeezing is reflected in
the vortex position distribution.
This is shown in Fig.~\ref{Dynamicsplot} where we also
monitor the time-evolution of  the deformation
$
\epsilon(t)={\rm max} \left[\frac{\langle \hat{x}_n^2(t)\rangle-
\langle \hat{y}_n^2(t)\rangle}
{\langle \hat{x}_n^2(t)\rangle+\langle \hat{y}_n^2(t)\rangle} \right]
$
of the vortex position distribution. Because the uncertainty ellipse that
characterizes the vortex position distribution precesses
in time, the deformation is maximized with respect
to the rotation angle of the coordinate system.

{\it Discussion.}--- We have developed a quantum theory of a vortex in
a one-dimensional optical lattice and applied the theory
to study the coupled quadrupole-Kelvin mode dynamics. Our solution of the Kelvin mode
dispersion relation essentially amounts to a solution of the vortex-line
dynamics in an optical lattice. Among other things our study revealed 
the possibility of squeezed vortex states and asymmetric
number distributions for the Kelvin modes.
As yet, there have been no experiments on vortices in optical lattices reported.
Detailed studies of the Kelvin mode dynamics would require imaging 
of the three-dimensional vortex line which is difficult, but
not impossible~\cite{Anderson2001a,Bretin2003a}. 
An alternative, but less 
direct way to probe the system, would be to track the evolution of the
quadrupole oscillations. If the resonance is narrow enough our theory predicts
revivals of these oscillations. In addition, the multimode nature of 
the Kelvin mode distribution 
can cause a decay rate of the quadrupole mode that is quite different from Fermi's
golden rule.
Our theory can be applied to a variety of phenomenon.
For example, the interplay between the superfluid Mott-insulator
transition, the quantum melting of 
a vortex lattice, and the quantum Hall regime in an optical lattice
is a very interesting topic for further research.

\vspace{-0.4cm}
This work is supported by the Stichting voor Fundamenteel Onderzoek der 
Materie (FOM) and by the Nederlandse Organisatie voor 
Wetenschaplijk Onderzoek (NWO).

\vspace{-0.7cm}

\end{document}